\documentclass[twocolumn,prl,showpacs]{revtex4}
\usepackage{epsf}

\newcommand{\Av}[1]     {\left\langle #1 \right\rangle}

\newcommand{\fig}[1]	{Fig.~\ref{#1}}

\def\bi         {\begin{itemize}}
\def\ei         {\end{itemize}}
\def\benu	{\begin{enumerate}}
\def\eenu	{\end{enumerate}}
\def\bmat       {\left[ \begin{array}}
\def\emat       {\end{array} \right]}
\def\beq	{\begin{equation}}
\def\eeq	{\end{equation}}
\def\beqn       {\begin{eqnarray*}}
\def\eeqn       {\end{eqnarray*}}
\def\beqa       {\begin{eqnarray}}
\def\eeqa       {\end{eqnarray}}
\def\bquote	{\begin{quote}}
\def\equote	{\end{quote}}
\def\f          {\frac}
\def\bwide	{\begin{widetext}}
\def\ewide	{\end{widetext}}

\def\d          {\delta}
\def\e          {\epsilon}

\def\m          {\mu}

\def\s          {\sigma}

\def\w          {\omega}

\def\z		{\zeta}

\def\D          {\Delta}

\def\bk         {{\bf k}}

\def\bq         {{\bf q}}

\def\hQ		{{\hat{Q}}}

\def\dag	{\dagger}

\def\tr		{{\mbox{Tr}~}}
\def\im		{{\mbox{Im}}}

\begin{document}

\title{A Direct Probe of Electronic Nematic Order;
Symmetry Information in Scanning Tunneling Microscope Images }
\author{Hyeonjin Doh}
%%\email{hdoh@physics.utoronto.ca}
%\affiliation{Department of Physics, University of Toronto, Toronto, 
%Ontario M5S 1A7 Canada}
\author{Hae-Young Kee}
%%\email{hykee@physics.utoronto.ca}
\affiliation{Department of Physics, University of Toronto, Toronto, 
Ontario M5S 1A7 Canada}
%\author{J. C. Davis}
%\affiliation{LASSP, Department of Physics, Cornell University, Ithaca, NY
%14853 USA}
\date{\today}
\begin{abstract}
An electronic nematic state spontaneously breaks a point-group symmetry of an underlying lattice.
As a result, the nematic-isotropic transition accompanies a Fermi surface distortion. 
However, the anisotropic nature of the nematic state at a macroscopic scale 
can be easily wiped out when domains of different orientations of nematic order exist.
We suggest that a spatial pattern of local density of states (LDOS) in the presence of 
a non-magnetic impurity can be a direct probe of the nematic order.
We study various patterns of LDOS across the quantum phase transition
between the isotropic and nematic phases.  Especially the Fourier transformed 
local density of states (FT-LDOS),
which can be deduced from scanning tunneling microscope  images,
represent a transparent symmetry of an electronic structure.
The application of our results to the bilayer ruthenate, Sr$_3$Ru$_2$O$_7$ is also discussed.
\end{abstract}

\maketitle

%\section{Introduction}
%{\it Introduction} --- 
\paragraph{Introduction:}
Recently, the existence of an intermediate form of matter dubbed the electronic nematic liquid
was proposed to explain novel features observed in ultra-pure Sr$_3$Ru$_2$O$_7$.\cite{Borzi07Science}
Especially, a large magnetoresistive anisotropy in the close vicinity of a metamagnetic quantum
critical point shows a striking similarity to those reported in
ultra-clean two-dimensional electron systems in high magnetic fields\cite{Lilly99prl,Du99ssc},
where the electronic nematic phase was also suggested.
The metamagnetic transition in Sr$_3$Ru$_2$O$_7$ possesses a critical end point at a finite
temperature, and it was proposed that
a quantum critical point can be achieved by tilting the magnetic field towards
the direction  perpendicular to the Ru-O plane.\cite{Grigera01Science}
However, when the system gets close to the putative quantum critical point, 
a new phase bounded by two successive metamagnetic transitions was found in
an ultra-clean sample.\cite{Grigera04Science} 
The emergent phase is characterized by a high residual resistivity compared to the phases nearby,
and it was suggested that
a formation of the electronic nematic phase leads to these remarkable phenomena of
consecutive metamagnetic transitions and the high residual resistivity.\cite{Grigera04Science,
KeeHY05prb,HDoh06}

The electronic nematic phase is characterized as an anisotropic metallic state
with a broken point-group symmetry of an underlying lattice.\cite{Yamase00jpsj,Halboth00prl,Kivelson03rmp,KeeHY03prb, Khavkine04prb, Yamase05prb} 
For example, on a square lattice, there are two
distinct nematic phases depending on how the
%90-degree
$\pi/2$
rotational symmetry is broken.\cite{DohH06prb}
A consequence of the broken symmetry is a spontaneous Fermi surface distortion, 
which leads to an anisotropy between two directions of the longitudinal
resistivity.
Indeed, a large magnetoresistive anisotropy was observed
in Sr$_3$Ru$_2$O$_7$ consistent with a formation of nematic phase.
On the other hand, when the field is perpendicular to the Ru-O plane,
the anisotropy is no longer present, but the  high residual resistivity becomes pronounced.
It was proposed that the nematic domain formation results in the 
high resistivity in the emergent phase, which involves two different orientations of 
the nematic order in the square lattice system.\cite{HDoh06}
%Fermi surface is elongated along the $x$-direction in parts of the sample, while 
%it is elongated along the $y$-direction in other domains.
In the presence of  the nematic domains,
the nature of the broken rotational symmetry can be hardly observable in any bulk measurement. 

In this paper, we suggest that a spatial pattern of the LDOS 
as measured by scanning tunneling microscopy (STM) provides a direct probe of the nematic order. 
Note that the nematic-isotropic transition is first order, thus the insignificant 
effect of order parameter fluctuation
on the Fermi surface measurement is in contrast to the case near the nematic quantum 
critical point.\cite{Lawler06prb}
We show that the Friedel oscillations around an impurity show
visible anisotropic patterns between $x$ and $y$ directions in the nematic phase, as one
expected.
% via the magnetic field.
% LDOS is insensitive to the nematic domain formation, 
%as long as a size of domain is bigger than a characteristic length scale associated
%with the nematic order.
In particular,  the FT-LDOS shows a transparent difference between the nematic and isotropic phases
across the transition by the magnetic field.
Therefore, STM on the bi-layer ruthenate is highly desirable
for the confirmation of the nematic phase emerging near the vicinity of the putative quantum
critical point.

%\section{}
%{\it Phenomenological Model} --
\paragraph{Phenomenological Model:}
%\label{sec:theory}
The essential physics of the broken symmetry possessed in LDOS is insensitive to 
the choice of a Hamiltonian,
as long as the nematic-isotropic transition is captured in the Hamiltonian.
On the other hand, the detailed patterns of the LDOS  depend on the interactions as
well as  band structure. 
To capture an electronic structure symmetry embedded in 
spatial patterns of LDOS across the transition, 
we use a phenomenological Hamiltonian with a quadrupole interaction\cite{Oganesyan01prb,KeeHY03prb},
which successfully described the phenomena of the metamagnetic transitions, 
high resistivity, and anisotropic magnetoresistivity in the bilayer ruthenate.\cite{KeeHY05prb,
HDoh06,Puetter07}  

\begin{figure}[htb]
\epsfxsize=7cm
\epsffile{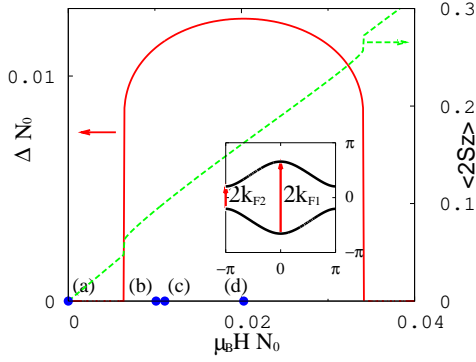}
\caption{Mean field result of nematic order (solid line)
and magnetization (dashed line) as a function of magnetic field.
The points marked as (a), (b), (c), and (d) are where we 
present LDOS in the following figures.
The inset shows the Fermi surface of up-spin and two dominant extreme
vectors in the nematic phase discussed in the main text.
\label{fig:meanfield}
}
\end{figure}

Since the nematic phase contains a broken rotational symmetry, an analog of the order
parameter of
the nematic liquid crystal is the first step towards the construction of a 
phenomenological model.\cite{Oganesyan01prb,KeeHY03prb,Khavkine04prb}
We defined $\hQ(\bq)$, a quadrupole tensor on  the 
square lattice as
%follows.
\beq
\hQ_\s(\bq) = \sum_\bk c_{\bk+\f{\bq}{2},\s}^\dag
\bmat{cc}
\z_1(\bk) & \z_2(\bk) \\
\z_2(\bk) & -\z_1(\bk)
\emat
c_{\bk-\f{\bq}{2},\s}^{},
\eeq
where $\z_1(\bk) = \cos k_x - \cos k_y$ and $\z_2(\bk) = 2\sin k_x \sin k_y$.
We can write down the Hamiltonian  as follows using the quadrupole tensor,
which leads to the nematic order within the mean-field theory. 
\beq
H =\! \sum_{\bk\s} (\e_\bk\!-\!\m-\!\s \m_B H)c_{\bk\s}^\dag c_{\bk\s}^{}
-\!\sum_{\bq\s} F_2(\bq) \tr\hQ_\s^\dag(\bq)\hQ_\s(\bq),
\label{eqn:nematic_hamil}
\eeq
where $\e_\bk = -t (\cos k_x + \cos k_y) $ is the electronic dispersion of  
the tight-binding model on the  square lattice,
and $\m$ is a chemical potential.
$\s\m_BB$ is a Zeeman term, where $\s = +1$ for up-spin and $\s=-1$ for down-spin.
$F_2(\bq)$ represents the strength of the quadrupole-quadrupole
interaction, which can be viewed as the 
Landu quasiparticle interaction of $l=2$ angular momentum channel.\cite{Pomeranchuk58jetp,Oganesyan01prb,MLawler06prb,Nilsson05prb,Quintanilla06prb}
%\beq
%F_2(\bq) = \f{F_2}{1+\k q^2}
%\eeq
%For the simplicity, we will consider the on-site interaction where $\k = 0$.

There are two types of the Fermi surface distortion associated with
$\z_1$ and $\z_2$ which represent the Fermi surface elongations along the directions 
parallel to the axis and  diagonal to the axis, respectively.
The previous mean-field results showed that
for a given value of $F_2$, the diagonal distortion does not occur,
so we focus on the nematic phase with a Fermi surface distorted along the $x$ or $y$-axis
of the square lattice.
The mean field equation for the uniform nematic order can be written as
\beq
H = \sum_{\bk\s} \left[\e_\bk - \D_\s \z_1(\bk) -\m -\s\m_BH\right]
c_{\bk\s}^\dag c_{\bk\s}^{}
+ \sum_\s\f{|\D_\s|^2}{2F_2},
\label{eqn:mean_hamiltonian}
\eeq
where the nematic order parameter is defined as
%\beq
$
\Delta_\s = \sum_\bk F_2(0) \z_1(\bk) \Av{c_{\bk\s}^\dag c_{\bk\s}^{}}.
$
%\label{eqn:nematic_order}
%\eeq
The order parameter as a function of magnetic field $H$ for a given chemical potential
was studied in Ref.\onlinecite{KeeHY05prb}.
The strongly first order nature of transition is originated from the van Hove singularity,
and the density of up(or down)-spin as a function of Zeeman field
jumps at the transitions, which results in two consecutive metamagnetic transitions.
In \fig{fig:meanfield}, we showed the earlier result of the order parameter to indicate 
the locations in the phase diagram where we present LDOS.
We set  $F_2 N_0=0.1$ and $\mu N_0 = -0.02$, where $N_0 = 2 \pi^2 t$.
(a) represents the isotropic phase with $\Delta N_0 =0$, and (b)-(d) the nematic phase with
$\Delta N_0 = 0.0113, 0.0116$ and $0.0126$, respectively.
We carried out the standard T-matrix formalism to compute FT-LDOS and
show the results for (a) -(d) cases below.

\begin{figure*}[hbt]
\epsfxsize=14cm
\epsffile{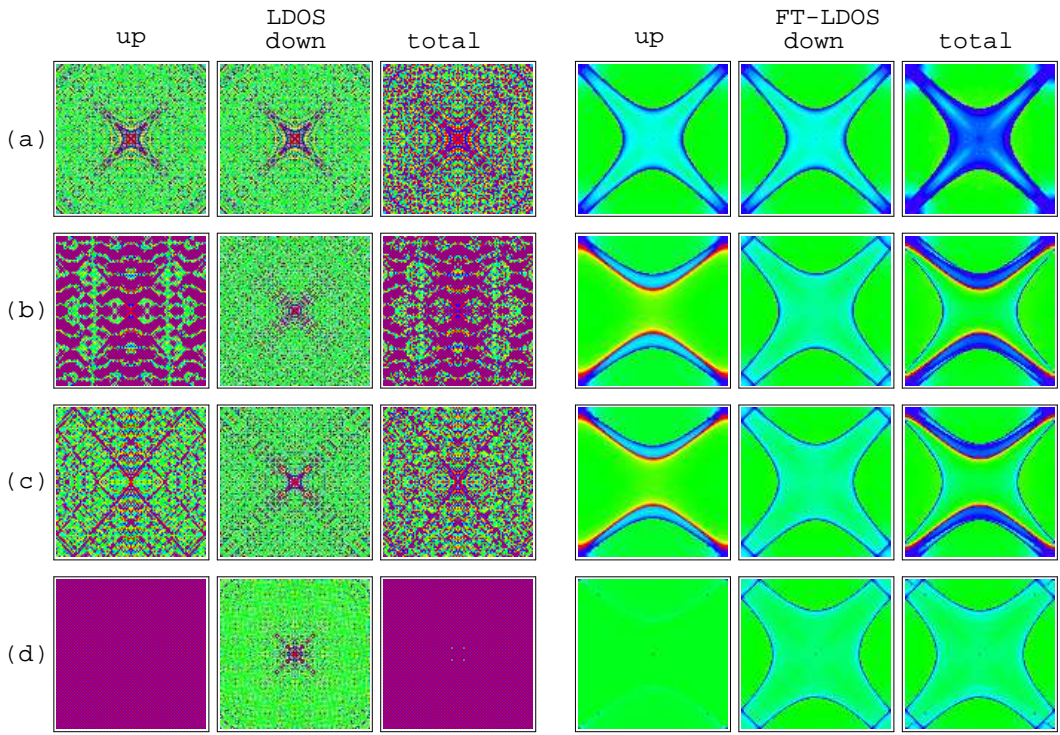}
\caption{LDOS and FT-LDOS at $\w = 0$ for the marked points, (a) - (d) in
\fig{fig:meanfield}. The left three columns are LDOS and
the right tree columns are FT-LDOS.
Each set of three columns shows spin-up, spin-down, and sum of both spins from left to right.
From top to bottom, nematic order increases as marked in \fig{fig:meanfield}.
\label{fig:ldos_array}
}
\end{figure*}

%{\it Local density of states  with a single impurity} --
\paragraph{Local density of states  with a single impurity:}
The FT-LDOS, $n(\bq;\w)$ with a single impurity is obtained 
using the standard T-matrix formalism.
%\cite{Mahan}
\beq
n(\bq;\w)
= n_0(\bq;\w)+\im \int d\bk 
 G_{\s}^{(0)}(\bk+\bq;\w)T_{\s\s'}(\w)G_{\s'}^{(0)}(\bk;\w),
\label{eqn:ldos_green}
\eeq
with the T-matrix,
%\beq
$
T^{-1}(\w) = 
V_s^{-1}  -  \int\!d\bk~G_{\s}^{(0)}(\bk;\w),
$
%\eeq
where $V_s$ represents the strength of the impurity, and we set $V_s N_0 = - 0.005$. 
The  single particle retarded Green's function, $G_{\s}^{(0)}$  
modified by the nematic order self-consistently can be written as follows.
\beq
G_{\s}^{(0)}
= \f{1}{\w - \left(\e_\bk -\D_\s\z_1(\bk) - \m - \s \m_BB\right) - i\d}.
\eeq
%where $\w$ corresponds to the bias voltage in STM measurement.
Here we ignore effects of the impurity on the nematic order,
so our results are valid only when the strength of an impurity and the bias voltage
is smaller than the energy scale of nematic order.
We use 2-dimensional adaptive integration to compute $n(\bq;w)$
 shown in \fig{fig:ldos_array}.

\begin{figure}[htb]
\epsfxsize=8.0cm
\epsffile{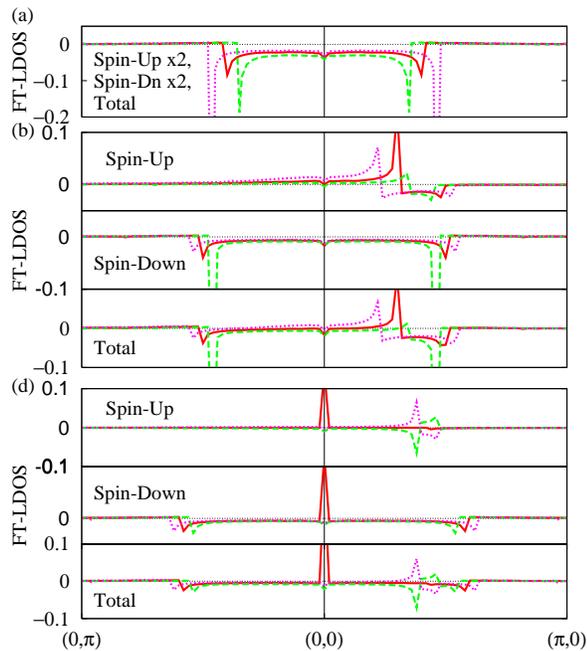}
\caption{FT-LDOS (subtracted by FT-LDOS in the absence of an impurity) 
along  $q_x$- and $q_y$-axis are shown for the marked points (a), (b), and (d) in \fig{fig:meanfield}.
The left-half is FT-LDOS along $q_y$-axis and the right-half, along $q_x$-axis.
The red solid line, the magenta dotted, and the green dashed lines correspond
to FT-LDOS at $\w=0$, $-0.1t$, and $0.1t$, respectively.
\label{fig:qaxis}
}
\end{figure}

\fig{fig:ldos_array} shows LDOS and FT-LDOS at the Fermi energy ($\w=0$)
for various strengths of nematic order corresponding to (a)-(d) indicated
in \fig{fig:meanfield}. 
The left three columns show
 LDOS in real space for up-spin, down-spin, and sum of both spins 
from left to right, respectively.
FT-LDOS images are also shown in the next three columns.
Each row corresponds to (a) to (d) cases in \fig{fig:meanfield} from top to bottom.
The color represents a relative strength of LDOS (FT-LDOS) induced by the
impurity.
The Friedel oscillations around the impurity located at the center show
visible anisotropic patterns between $x$ and $y$ directions.  
The peak structures in FT-LDOS are associated with the Fermi surface nesting
vectors. In the isotropic phase of (a), the peaks of the FT-LDOS form
an asteroidal shape with the strongest peak near ($\pm\pi$,$\pm\pi$) which is the 
nesting vector of the half-filled tight-binding dispersion.
In the nematic states, (b) and (c), the asteroidal peak structure abruptly changes to
a sandglass-like shape due to the opening of the Fermi surface along the $x$-axis.
The anisotropic shape in the nematic phase is mainly due to the extreme vectors of 
$2\bk_{F1}$  and $2\bk_{F2}$ as shown in the inset of \fig{fig:meanfield}.
These extreme vectors also determine the wave length of the Friedel oscillations
which will be further discussed below.
While the nematic order is maximized in (d), we found that both FT-LDOS and LDOS do not
show a significant signal of the nematic phase because of the destructive
interference between two extreme vectors of $2\bk_{F2}$ and $2 \bk_{F1}$.
However, the nematic order signals at finite bias voltages represented 
by the dashed and dotted lines in the bottom of \fig{fig:qaxis}.

\begin{figure}[htb]
\epsfxsize=8.0cm
\epsffile{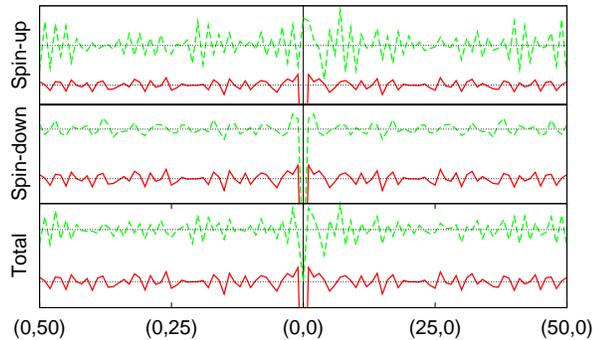}
\caption{LDOS for $\w=0$ are
shown along $x$- and $y$- axis with an impurity at the origin.
The left half of the figure shows LDOS along $y$-axis, and
the right half, along $x$-axis.
The solid and dashed lines correspond to $\Delta N_0  =0$ and $0.0116$, respectively.
\label{fig:raxis}
}
\end{figure}

FT-LDOS  along $q_x$- and $q_y$-axis  for finite $\w$
are shown in \fig{fig:qaxis}.
The  dramatic changes in FT-LDOS occur across the transition. 
FT-LDOS along the $q_x$ direction lose their weight, while
additional structures develop along the $q_y$ direction.
One is  originated from the extreme vector of $2\bk_{F2}$ connecting $(\pi,\pm k_{F2})$,
while the other from the extreme vector of $2\bk_{F1}$ connecting $(0, \pm k_{F1})$,
of the Fermi surface shown in \fig{fig:meanfield}.
These peaks along the $q_y$ direction give a unique pattern of Friedel
oscillation in  LDOS.
LDOS along $x$ and $y$-axis shown in \fig{fig:raxis} also 
reflect the broken symmetry in the nematic state.
For example, LDOS represented by the dashed line clearly show different amplitude variations 
along the $y$-direction, which 
roughly correspond to the wave length $(k_{F1}-k_{F2})^{-1}$.

\begin{figure}[htb]
\epsfxsize=8.0cm
\epsfysize=5.0cm
\epsffile{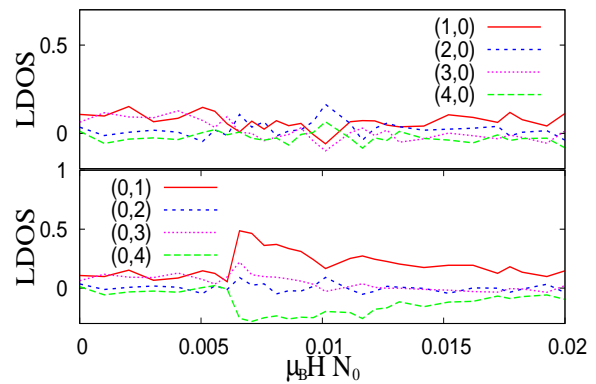}
\caption{LDOS (at $\w=0$) vs. magnetic field at several neighboring sites
(with lattice spacing $a \equiv 1$) around the impurity.
\label{fig:Bscan}
}
\end{figure}

The magnetic field dependence of LDOS is shown in \fig{fig:Bscan}.
At the isotropic-nematic transition occurring around $\mu_B H = 0.13 t$, LDOS at impurity site
shows a sudden decrease (increase) for an attractive (repulsive) impurity potential
(not shown here).
It is interesting note that LDOS at neighboring sites such as $(0,1)$ (in unit of
the lattice spacing $a \equiv 1$), show a dramatic change across the transition,
while LDOS at $(1,0)$ does not show any significant change across the transition.
The field dependence of LDOS also possesses the strong anisotropic nature of
the nematic phase.

%{\it Discussion and Summary} --
\paragraph{Discussion and Summary:}
%\label{sec:conclude}
Strongly correlated electron systems with $d$ or $f$-electrons
%like heavy fermions or high-T$_c$ superconductors 
have demonstrated rich phase diagrams with many different phases.
When the transition temperature between two different phases is driven to zero by 
the application of a magnetic field,  pressure,  or via chemical doping, we
can achieve the quantum critical point (QCP).
%, where the transition temperature is zero.  
The behavior of the system close to the QCP is very different 
from the behavior in the vicinity of the classical phase transition, and it has been an 
intensive research area.
Along the journey of searching for QCP, we often discovered that a novel phase
emerges at the vicinity of the desired QCP.
%\cite{HF, Organic_S}
%This has happened in a ultra-clean sample of the bilayer ruthenate. 

The emergent phase at the vicinity of the QCP in the bilayer ruthenate
is characterized by the anisotropic magnetoresistivity, when the field is
tilted from c-axis. 
On the other hand, the high resistivity without anisotropy was reported, 
when the field is along c-axis, $H \parallel c$. 
It was proposed that the nematic domain formation is responsible for the high resistivity
for $H \parallel c$. These imply
that the nematic domain may align when the field is tilted
from c-axis mostly due to the in-plane component of magnetic field,
and results in the anisotropic magnetoresistivity.
At present, we do not have a microscopic picture which describes
interplay between the nematic domain formation and the direction of magnetic field,
which needs to be further investigated in future.

Here we proposed that a local probe such as STM is insensitive to a macroscopic inhomogeneity,
and therefore it can directly confirm an existence of the nematic order.
We studied  FT-LDOS and LDOS in the presence of a nonmagnetic impurity 
across the nematic-isotropic transition tuned by the external magnetic field,
and found the followings.
(1)  FT-LDOS  as well as the Friedel oscillations in real space 
show a transparent indication of the broken rotational symmetry.
In particular, the Fourier transformed image of LDOS which can be deduced from
STM data gives direct information on the topology of a Fermi surface
associated with the electronic nematic order.
(2) Across the transition,
the change of the Fermi surface volume accompanies a sudden change
in LDOS. Therefore, LDOS at the impurity and neighboring sites 
show significant and anisotropic changes  across the transition.
(3) In the nematic phase, it is possible that the anisotropic signal can be washed out,
when the signals from two extreme vectors have destructive interference. 
However, anisotropic properties are still detectable at finite $\w$. 
%so one should examine the FT-LDOS at finite bias voltage in STM.
%\ei
%

\begin{acknowledgments}
We thank  J. C. Davis for encouraging us to present the current work.
This work was supported by NSERC of Canada, Canada Research
Chair, Canadian Institute for Advanced Research, and Alfred P.~Sloan Foundation.
\end{acknowledgments}


\begin{thebibliography}{13}
\expandafter\ifx\csname natexlab\endcsname\relax\def\natexlab#1{#1}\fi
\expandafter\ifx\csname bibnamefont\endcsname\relax
  \def\bibnamefont#1{#1}\fi
\expandafter\ifx\csname bibfnamefont\endcsname\relax
  \def\bibfnamefont#1{#1}\fi
\expandafter\ifx\csname citenamefont\endcsname\relax
  \def\citenamefont#1{#1}\fi
\expandafter\ifx\csname url\endcsname\relax
  \def\url#1{\texttt{#1}}\fi
\expandafter\ifx\csname urlprefix\endcsname\relax\def\urlprefix{URL }\fi
\providecommand{\bibinfo}[2]{#2}
\providecommand{\eprint}[2][]{\url{#2}}

\bibitem[{\citenamefont{Borzi et~al.}(2007)\citenamefont{Borzi, Grigera,
  Farrell, Perry, Lister, Lee, Tennant, Maeno, and Mackenzie}}]{Borzi07Science}
\bibinfo{author}{\bibfnamefont{R.~A.} \bibnamefont{Borzi}},
  \bibinfo{author}{\bibfnamefont{S.~A.} \bibnamefont{Grigera}},
  \bibinfo{author}{\bibfnamefont{J.}~\bibnamefont{Farrell}},
%  \bibnamefont{et~al.},
  \bibinfo{author}{\bibfnamefont{R.~S.} \bibnamefont{Perry}},
  \bibinfo{author}{\bibfnamefont{S.~J.~S.} \bibnamefont{Lister}},
  \bibinfo{author}{\bibfnamefont{S.~L.} \bibnamefont{Lee}},
  \bibinfo{author}{\bibfnamefont{D.~A.} \bibnamefont{Tennant}},
  \bibinfo{author}{\bibfnamefont{Y.}~\bibnamefont{Maeno}}, \bibnamefont{and}
  \bibinfo{author}{\bibfnamefont{A.~P.} \bibnamefont{Mackenzie}},
  \bibinfo{journal}{Science} \textbf{\bibinfo{volume}{315}},
  \bibinfo{pages}{214} (\bibinfo{year}{2007}).

\bibitem[{\citenamefont{Lilly et~al.}(1999)\citenamefont{Lilly, Cooper,
  Eisenstein, Pfeiffer, and West}}]{Lilly99prl}
\bibinfo{author}{\bibfnamefont{M.~P.} \bibnamefont{Lilly}},
  \bibinfo{author}{\bibfnamefont{K.~B.} \bibnamefont{Cooper}},
  \bibinfo{author}{\bibfnamefont{J.~P.} \bibnamefont{Eisenstein}},
%  \bibnamefont{et~al.},
  \bibinfo{author}{\bibfnamefont{L.~N.} \bibnamefont{Pfeiffer}},
  \bibnamefont{and} \bibinfo{author}{\bibfnamefont{K.~W.} \bibnamefont{West}},
  \bibinfo{journal}{Phys. Rev. Lett.} \textbf{\bibinfo{volume}{82}},
  \bibinfo{pages}{394} (\bibinfo{year}{1999}).

\bibitem[{\citenamefont{Du et~al.}(1999)\citenamefont{Du, Tsui, Stormer,
  Pfeiffer, Baldwin, and West}}]{Du99ssc}
\bibinfo{author}{\bibfnamefont{R. ~R} \bibnamefont{Du}},
  \bibinfo{author}{\bibfnamefont{D. ~C.} \bibnamefont{Tsui}},
  \bibinfo{author}{\bibfnamefont{H.~L.} \bibnamefont{Stormer}},
  \bibinfo{author}{\bibfnamefont{L.~N.} \bibnamefont{Pfeiffer}},
  \bibinfo{author}{\bibfnamefont{K.~W.} \bibnamefont{Baldwin}},
  \bibnamefont{and} \bibinfo{author}{\bibfnamefont{K.~W.} \bibnamefont{West}},
  \bibinfo{journal}{Solid State Comm.} \textbf{\bibinfo{volume}{109}},
  \bibinfo{pages}{389} (\bibinfo{year}{1999}).


\bibitem[{\citenamefont{Grigera et~al.}(2001)\citenamefont{Grigera, Perry,
  Schofield, Chiao, Julian, Lonzarich, Ikeda, Maeno, Millis, and
  Mackenzie}}]{Grigera01Science}
\bibinfo{author}{\bibfnamefont{S.~A.} \bibnamefont{Grigera}},
  \bibinfo{author}{\bibfnamefont{R.~S.} \bibnamefont{Perry}},
  \bibinfo{author}{\bibfnamefont{A.~J.} \bibnamefont{Schofield}},
%  \bibnamefont{et~al.},
  \bibinfo{author}{\bibfnamefont{M.}~\bibnamefont{Chiao}},
  \bibinfo{author}{\bibfnamefont{S.~R.} \bibnamefont{Julian}},
  \bibinfo{author}{\bibfnamefont{G.~G.} \bibnamefont{Lonzarich}},
  \bibinfo{author}{\bibfnamefont{S.~I.} \bibnamefont{Ikeda}},
  \bibinfo{author}{\bibfnamefont{Y.}~\bibnamefont{Maeno}},
  \bibinfo{author}{\bibfnamefont{A.~J.} \bibnamefont{Millis}},
  \bibnamefont{and} \bibinfo{author}{\bibfnamefont{A.~P.}
  \bibnamefont{Mackenzie}},
\bibinfo{journal}{Science}
  \textbf{\bibinfo{volume}{294}}, \bibinfo{pages}{329} (\bibinfo{year}{2001}).

\bibitem[{\citenamefont{Grigera et~al.}(2004)\citenamefont{Grigera, Gegenwart,
  Borzi, Weickert, Schofield, Perry, Tayama, Sakakibara, Maeno, Green
  et~al.}}]{Grigera04Science}
\bibinfo{author}{\bibfnamefont{S.~A.} \bibnamefont{Grigera}},
  \bibinfo{author}{\bibfnamefont{P.}~\bibnamefont{Gegenwart}},
  \bibinfo{author}{\bibfnamefont{R.~A.} \bibnamefont{Borzi}},
%  \bibnamefont{et~al.},
  \bibinfo{author}{\bibfnamefont{F.}~\bibnamefont{Weickert}},
  \bibinfo{author}{\bibfnamefont{A.~J.} \bibnamefont{Schofield}},
  \bibinfo{author}{\bibfnamefont{R.~S.} \bibnamefont{Perry}},
  \bibinfo{author}{\bibfnamefont{T.}~\bibnamefont{Tayama}},
  \bibinfo{author}{\bibfnamefont{T.}~\bibnamefont{Sakakibara}},
  \bibinfo{author}{\bibfnamefont{Y.}~\bibnamefont{Maeno}},
  \bibinfo{author}{\bibfnamefont{A.~G.} \bibnamefont{Green}},
  \bibinfo{journal}{Science}
  \textbf{\bibinfo{volume}{306}}, \bibinfo{pages}{1154} (\bibinfo{year}{2004}).

\bibitem[{\citenamefont{Kee and Kim}(2005)}]{KeeHY05prb}
\bibinfo{author}{\bibfnamefont{H.-Y.} \bibnamefont{Kee}} \bibnamefont{and}
  \bibinfo{author}{\bibfnamefont{Y.~B.} \bibnamefont{Kim}},
  \bibinfo{journal}{Phys. Rev. B} \textbf{\bibinfo{volume}{71}},
  \bibinfo{pages}{184402} (\bibinfo{year}{2005}).

\bibitem[{\citenamefont{Doh et~al.}(2006{\natexlab{b}})\citenamefont{Doh, Kim,
  and Ahn}}]{HDoh06}
\bibinfo{author}{\bibfnamefont{H.}~\bibnamefont{Doh}},
  \bibinfo{author}{\bibfnamefont{Y.~B.} \bibnamefont{Kim}}, \bibnamefont{and}
  \bibinfo{author}{\bibfnamefont{K.~H.} \bibnamefont{Ahn}}
  (\bibinfo{year}{2006}{\natexlab{b}}), \bibinfo{note}{cond-mat/0604425}.

\bibitem[{\citenamefont{Yamase et~al.}(2000{\natexlab{a}})\citenamefont{Yamase, and Kohno}}]{Yamase00jpsj}
\bibinfo{author}{\bibfnamefont{H.}~\bibnamefont{Yamase}},
\bibnamefont{and}
  \bibinfo{author}{\bibfnamefont{H.} \bibnamefont{Kohno}}
  \bibinfo{journal}{J. Phys. Soc. Jpn.} \textbf{\bibinfo{volume}{69}},
  \bibinfo{pages}{332} (\bibinfo{year}{2000}).
% ; \textbf{\bibinfo{volume}{69}
%\bibinfo{pages}{2151} (\bibinfo{year}{2000}).

\bibitem[{\citenamefont{Halboth and Metzner}(2000)}]{Halboth00prl}
\bibinfo{author}{\bibfnamefont{C.~J.} \bibnamefont{Halboth}} \bibnamefont{and}
  \bibinfo{author}{\bibfnamefont{W.}~\bibnamefont{Metzner}},
  \bibinfo{journal}{\prl} \textbf{\bibinfo{volume}{85}}, \bibinfo{pages}{5162}
  (\bibinfo{year}{2000}).

\bibitem[{\citenamefont{Kivelson et~al.}(2003)\citenamefont{Kivelson, Fradkin,
  Oganesyan, Bindloss, Tranquada, Kapitulnik, and Howald}}]{Kivelson03rmp}
\bibinfo{author}{\bibfnamefont{S.~A.} \bibnamefont{Kivelson}},
  \bibinfo{author}{\bibfnamefont{E.}~\bibnamefont{Fradkin}},
  \bibinfo{author}{\bibfnamefont{V.}~\bibnamefont{Oganesyan}},
  \bibinfo{author}{\bibfnamefont{I.~P.} \bibnamefont{Bindloss}},
  \bibinfo{author}{\bibfnamefont{J.~M.} \bibnamefont{Tranquada}},
  \bibinfo{author}{\bibfnamefont{A.}~\bibnamefont{Kapitulnik}},
  \bibnamefont{and} \bibinfo{author}{\bibfnamefont{C.}~\bibnamefont{Howald}},
  \bibinfo{journal}{Rev. Mod. Phys.} \textbf{\bibinfo{volume}{75}},
  \bibinfo{pages}{1201} (\bibinfo{year}{2003}).

\bibitem[{\citenamefont{Kee et~al.}(2003)\citenamefont{Kee, Kim, and
  Chung}}]{KeeHY03prb}
\bibinfo{author}{\bibfnamefont{H.-Y.} \bibnamefont{Kee}},
  \bibinfo{author}{\bibfnamefont{E.~H.} \bibnamefont{Kim}}, \bibnamefont{and}
  \bibinfo{author}{\bibfnamefont{C.-H.} \bibnamefont{Chung}},
  \bibinfo{journal}{Phys. Rev. B} \textbf{\bibinfo{volume}{68}},
  \bibinfo{pages}{245109} (\bibinfo{year}{2003}).

\bibitem[{\citenamefont{Khavkine et~al.}(2004)\citenamefont{Khavkine, Chung,
  Oganesyan, and Kee}}]{Khavkine04prb}
\bibinfo{author}{\bibfnamefont{I.}~\bibnamefont{Khavkine}},
  \bibinfo{author}{\bibfnamefont{C.-H.} \bibnamefont{Chung}},
  \bibinfo{author}{\bibfnamefont{V.}~\bibnamefont{Oganesyan}},
%  \bibinfo{author}{\bibfnamefont{et~al.}},
  \bibnamefont{and} \bibinfo{author}{\bibfnamefont{H.-Y.} \bibnamefont{Kee}},
  \bibinfo{journal}{Phys. Rev. B} \textbf{\bibinfo{volume}{70}},
  \bibinfo{pages}{155110} (\bibinfo{year}{2004}).


\bibitem[{\citenamefont{Yamase et~al.}(2005{\natexlab{a}})\citenamefont{Yamase, Oganesyan, and Metzner}}]{Yamase05prb}
\bibinfo{author}{\bibfnamefont{H.}~\bibnamefont{Yamase}},
  \bibinfo{author}{\bibfnamefont{V.}~\bibnamefont{Oganesyan}}, \bibnamefont{and}  \bibinfo{author}{\bibfnamefont{W.} \bibnamefont{Metzner}}
  \bibinfo{journal}{Phys. Rev. B} \textbf{\bibinfo{volume}{72}},
  \bibinfo{pages}{35114} (\bibinfo{year}{2005}).

\bibitem[{\citenamefont{Doh et~al.}(2006{\natexlab{a}})\citenamefont{Doh,
  Friedman, and Kee}}]{DohH06prb}
\bibinfo{author}{\bibfnamefont{H.}~\bibnamefont{Doh}},
  \bibinfo{author}{\bibfnamefont{N.}~\bibnamefont{Friedman}}, \bibnamefont{and}
  \bibinfo{author}{\bibfnamefont{H.-Y.} \bibnamefont{Kee}},
  \bibinfo{journal}{Phys. Rev. B} \textbf{\bibinfo{volume}{73}},
  \bibinfo{pages}{125117} (\bibinfo{year}{2006}{\natexlab{a}}).

\bibitem[{\citenamefont{Lawler et~al.}(2006{\natexlab{a}})\citenamefont{Lawler,
  , and Fradkin}}]{Lawler06prb}
\bibinfo{author}{\bibfnamefont{M.}~\bibnamefont{Lawler}},
\bibnamefont{and}
  \bibinfo{author}{\bibfnamefont{E.} \bibnamefont{Fradkin}},
  \bibinfo{journal}{Phys. Rev. B} \textbf{\bibinfo{volume}{75}},
  \bibinfo{pages}{033304} (\bibinfo{year}{2006}{\natexlab{a}}).


\bibitem[{\citenamefont{Oganesyan et~al.}(2001)\citenamefont{Oganesyan,
  Kivelson, and Fradkin}}]{Oganesyan01prb}
\bibinfo{author}{\bibfnamefont{V.}~\bibnamefont{Oganesyan}},
  \bibinfo{author}{\bibfnamefont{S.~A.} \bibnamefont{Kivelson}},
  \bibnamefont{and} \bibinfo{author}{\bibfnamefont{E.}~\bibnamefont{Fradkin}},
  \bibinfo{journal}{Phys. Rev. B} \textbf{\bibinfo{volume}{64}},
  \bibinfo{pages}{195109} (\bibinfo{year}{2001}).

\bibitem[{\citenamefont{Puetter et~al.}(2007)\citenamefont{Puetter, Doh, and
  Kee}}]{Puetter07}
\bibinfo{author}{\bibfnamefont{C.}~\bibnamefont{Puetter}},
  \bibinfo{author}{\bibfnamefont{H.}~\bibnamefont{Doh}}, \bibnamefont{and}
  \bibinfo{author}{\bibfnamefont{H.-Y.} \bibnamefont{Kee}}
  (\bibinfo{year}{2007}), \bibinfo{note}{unpublished}.

\bibitem[{\citenamefont{Pomeranchuk}(1958)}]{Pomeranchuk58jetp}
\bibinfo{author}{\bibfnamefont{I.~J.} \bibnamefont{Pomeranchuk}},
  \bibinfo{journal}{Sov. Phys. JETP} \textbf{\bibinfo{volume}{8}},
  \bibinfo{pages}{361} (\bibinfo{year}{1958}).

\bibitem[{\citenamefont{Lawler et~al.}(2006{\natexlab{a}})\citenamefont{Lawler,
  , and Fradkin}}]{MLawler06prb}
\bibinfo{author}{\bibfnamefont{M.} \bibnamefont{Lawler}},
\bibinfo{author}{\bibfnamefont{V.} \bibnamefont{Fernandez}},
\bibinfo{author}{\bibfnamefont{D.} \bibnamefont{Barci}},
  \bibinfo{author}{\bibfnamefont{E.} \bibnamefont{Fradkin}},
\bibnamefont{and}
  \bibinfo{author}{\bibfnamefont{L.} \bibnamefont{Oxman}},
  \bibinfo{journal}{Phys. Rev. B} \textbf{\bibinfo{volume}{73}},
  \bibinfo{pages}{085101} (\bibinfo{year}{2006}{\natexlab{a}}).


\bibitem[{\citenamefont{Nilsson and Neto}(2005)}]{Nilsson05prb}
\bibinfo{author}{\bibfnamefont{J.}~\bibnamefont{Nilsson}} \bibnamefont{and}
  \bibinfo{author}{\bibfnamefont{A.~H.~C.} \bibnamefont{Neto}},
  \bibinfo{journal}{\prb} \textbf{\bibinfo{volume}{72}}, \bibinfo{eid}{195104}
  (\bibinfo{year}{2005}).


\bibitem[{\citenamefont{Quintanilla and Schofield}(2006)}]{Quintanilla06prb}
\bibinfo{author}{\bibfnamefont{J.}~\bibnamefont{Quintanilla}} \bibnamefont{and}
  \bibinfo{author}{\bibfnamefont{A. ~J.} \bibnamefont{Schofield}},
  \bibinfo{journal}{\prb} \textbf{\bibinfo{volume}{74}}, \bibinfo{eid}{115126}
  (\bibinfo{year}{2006}).


\end{thebibliography}
\end{document}